\title{Maximum Eigenmode Relaying with statistical Channel State Information 
at the Relay}
\documentclass[9pt,twocolumn,balance]{IEEEtran}
\usepackage{graphicx,subfigure,balance,amsthm,amsmath,amssymb,color,url,setspace,cite}
\usepackage[countmax]{subfloat}
\graphicspath{{./figs/}}
\DeclareGraphicsExtensions{{.eps}{.pdf}{.jpeg}}

\author{\IEEEauthorblockN{Mehdi Molu, Norbert Goertz\\}
\IEEEauthorblockA{Institute of Telecommunications,
Vienna University of Technology
}}

\begin{document}
\maketitle
\IEEEpeerreviewmaketitle
\begin{abstract}
Optimal precoding in the relay is investigated to maximize ergodic
capacity of a multiple antenna relay channel. The source and the relay
nodes are equipped with multiple antennas and the destination with a
single antenna. It is assumed that the channel covariance matrices of
the relay's receive and transmit channels are available to the relay, and
optimal precoding at the relay is investigated. It is shown that the
optimal transmission from the relay should be conducted in the
direction of the eigenvectors of the transmit-channel covariance
matrix. Then, we derive the necessary and sufficient conditions under
which the relay transmission only from the strongest eigenvector 
achieves capacity; this method is called Maximum Eigenmode Relaying
(MER).


\end{abstract}
\newtheorem{thm}{Theorem}
\newtheorem{cor}{Corollary}
\newtheorem{lem}{Lemma}
 \section{Introduction and Related Work}
\label{Sec:Introduction}
Cooperative communication has been a rather active field of research
 in the recent past
 (e.g. \cite{LaTsWo:2004,SeErAa-Part1:2003,SeErAa-Part2:2003,CoGa:1979}).
 It was first shown by van der Meulen in \cite{Va:1971} that
 cooperation can enhance the transmission rate of a communication
 system. Later on, substantial work was carried out, investigating the
 effect of cooperation in various types of communication systems. The
 first works concentrated on communication nodes with single antennas
 and several relaying protocols were proposed of which two important
 ones are (non-regenerative) AF (Amplify-and-Forward) relaying and
 (regenerative) DF (Decode-and-Forward) relaying. In this paper, we will
 focus on non-regenerative relaying techniques.

Multiple antenna systems are well known to boost the Shannon capacity
of a communication system (see e.g.~\cite{Te:1995}). A natural setup
is the combination of cooperative communications with multiple
antennas, which is at the heart of current research in the
field. Assuming multiple antennas at the relay, one of the major tasks
is to design a suitable relaying protocol. However, depending on the
system or the desired performance criterion, different ``optimal''
relaying protocols can exist: for instance, the non-regenerative
relaying protocols, e.g.~in \cite{GuHa:2008,MoCh:2009,KhRo:2010}, are
designed to minimize Mean Square Error (MSE) but other relaying
protocols,
e.g.~in \cite{MuViAg:2007,TaHu:2007,DhMcMaBe:2011,JeSeLeKiKi:2012},
are assumed to maximize Shannon capacity. Moreover, depending on the
available channel information in the relay, the ``optimal'' relaying
protocol can be different.

\subsection{Related Work} \label{Sec:Related Work} 
A noncoherent cooperative system was investigated in \cite{DhMcMaBe:2011}.
  The source and the relay nodes are equipped with multiple antennas
and the destination node with single antenna. 
It was assumed that the source and the relay
have access to the covariance matrix of the channels. Another
assumption was that the antennas in the source are correlated but no
correlation was assumed in the relay. Using that knowledge, the
optimal transmit direction in the source and the relay was
derived. Assume $\boldsymbol{Q}$ is transmit covariance matrix in the
source with spectral decomposition $\boldsymbol{Q}
= \boldsymbol{U}_{Q} \boldsymbol{\Lambda}_{Q} \boldsymbol{U}_{Q}^H $
and $\boldsymbol{\Sigma}$ is the correlation matrix of the source,
with spectral decomposition $\boldsymbol{\Sigma}
= \boldsymbol{U}_\Sigma \boldsymbol{\Lambda}_\Sigma \boldsymbol{U}_\Sigma^H
$. It was proved that the optimal $\boldsymbol{Q}$ must be in the
form of $\boldsymbol{Q}^o
= \boldsymbol{U}_\Sigma \boldsymbol{\Lambda}_{Q^o} \boldsymbol{U}_\Sigma^H
$; i.e. $ \boldsymbol{U}_{Q^o}= \boldsymbol{U_\Sigma}$ and
$\boldsymbol{\Lambda}_{Q^o}$ is in descending order which needs to be
solved numerically. It was also proved that the optimal gain matrix
in the relay is a weighted identity matrix which depends on
$n_{\text{R}}$, $\boldsymbol{\Sigma}$ and
$\boldsymbol{Q}$. Furthermore, the necessary and sufficient condition
for the optimality of beamforming in the source was derived
as
\begin{eqnarray} \label{eq:RelatedWork--Prathapa} \gamma \lambda^\Sigma_2 \le \frac{(1+\gamma \lambda^\Sigma_1)\mathcal{D}(1+\gamma \lambda^\Sigma_1)}{\mathcal{D}(1+\gamma \lambda^\Sigma_1)+\mathcal{A}(\gamma \lambda^\Sigma_1)-1}-1 \end{eqnarray}
where $\mathcal{D}(1+\gamma \lambda^\Sigma_1)$ is a function given
in \cite[Eq. 16]{DhMcMaBe:2011} and
$\mathcal{A}(\gamma \lambda^\Sigma_1) = \mathbb{E}\left\lbrace (1+z)
e^{1+z}\Gamma(0,z) \right\rbrace$ where $\mathbb{E}\left\lbrace\cdot\right\rbrace$ represents expectation operation. 
Note that the evaluation of the optimality of beamforming in (\ref{eq:RelatedWork--Prathapa})
requires computationally expensive monte carlo simulations  or numerical integrations due to
$\mathcal{A}(\gamma \lambda^\Sigma_1)$.

 More recent results
were obtained in \cite{JeSeLeKiKi:2012} where an optimal relay
precoding was investigated for a system with correlated antennas at
the relay. It was assumed that the relay has access to full CSI of
$\boldsymbol{H}_1$ but only to the covariance matrix of
$\boldsymbol{H}_2$, i.e. to $\boldsymbol{R}_\text{R}$ with spectral
decomposition $\boldsymbol{R}_\text{R}
= \boldsymbol{U}_\text{R} \boldsymbol{\Lambda}_\text{R} \boldsymbol{U}_\text{R}^H
$. In \cite{JeSeLeKiKi:2012,JeKi:2009}, it was proved that the
optimal relay precoding matrix, $\boldsymbol{F}^o$, is in the form of
$\boldsymbol{F}^o
= \boldsymbol{U}_\text{R} \boldsymbol{\Lambda}^o_{F}\boldsymbol{V}^H_{H_1}$,
where $\boldsymbol{V}_{H_1}$ is the unitary matrix with columns as
the eigenvectors of $\boldsymbol{H}_1\boldsymbol{H}_1^H$. Along with
the numerical methods to derive $\boldsymbol{\Lambda}^o_{F}$
in \cite{JeSeLeKiKi:2012} , the optimality of beamforming was
considered \textit{only} for the asymptotic case of high transmit SNR. The
relay beamforming was found to be optimal if following inequality
holds: 
 \begin{eqnarray} \label{eq:RelatedWork--Cheol}
&&\hspace{-15mm}(\frac{\sigma^2}{P_\text{R} \lambda_1^\text{R}})^{n_\text{D}}
e^{\sigma^2/P_\text{R} \lambda_1^\text{R}} \Gamma(1-n_\text{D},\frac{\sigma^2}{P_\text{R} \lambda_1^\text{R}})\nonumber\\
&&\times
( \frac{P_\text{R} \lambda_2^\text{R}}{\sigma^2}+1)+ \frac{P_\text{R} \lambda_2^\text{R}}{\sigma^2}(n_\text{D}-1) \le 1
 \end{eqnarray}
Details can be found in \cite[Sec. III-C]{JeSeLeKiKi:2012}.

In this paper, we consider a MIMO cooperative system with multiple
antennas at the relay in which the relay has access to the covariance
matrices of preceding and following channels. Moreover, the relay
antennas are assumed to be correlated. With partial channel knowledge
(the covariance matrices) available to the relay, a capacity
maximizing relaying method will be introduced; the exact system model
and the desired performance criterion will be discussed in forthcoming
sections.

The paper is organized as follows: In Section \ref{Sec:System Model
and Problem Statement}, the system model is explained and two problems
are stated that are the main subject of this
paper. Section \ref{Sec:Optimal Precoding in the Relay} deals with
designing optimal precoding in the relay in order to maximise
capacity. In Section \ref{Sec:Optimality of Maximum Eigenmode
Relaying}, the necessary and sufficient conditions for the optimality
of MER are investigated. In Section \ref{Sec:Discussion} some
asymptotic results and their usefulness in practical systems are
considered. In Section \ref{Sec:Numerical Results} numerical results
are presented and finally, some conclusion remarks are explained in Section \ref{Sec:Conclusion}. 

\section{System Model and Problem Statement}
\label{Sec:System Model and Problem Statement}

\subsection{Notation}
\label{Subsec:Notation}
Matrices are represented by boldface upper cases
($\boldsymbol{H}$). Column and row vectors are denoted by boldface
lower cases ($\boldsymbol{h}$), and $h_i$ indicates the $i$-th element
of $\boldsymbol{h}$. The superscript $^H$ stands for Hermitian
transposition. We refer to the identity matrix by
$\boldsymbol{I}$. The expectation operation is indicated by
$\mathbb{E}(\cdot)$ and $f_X(x)$ is reserved for probability density
functions (pdf); $\boldsymbol{\Lambda}_\Sigma$ represents a diagonal
matrix with elements organized in descending order and
$\lambda_i^{\Sigma}$ denotes the $i$-th diagonal element of
$\boldsymbol{\Lambda}_\Sigma$. For simplicity of notation,
$(\lambda_i^{\Sigma})^2$ is abbreviated by $\lambda_i^{\Sigma2}$. The
trace of a matrix is denoted by $\mathrm{Tr}(\cdot)$.

\subsection{System Model}
\label{Subsec:System Model}
A dual hop, half duplex non-coherent MIMO communication system is
considered in this paper. A source node with $n_\text{S}$ antennas
communicates with a single-antenna destination node \emph{only} via a
relay node that is equipped with $n_\text{R}$ antennas (that are used
for both reception and transmission). It is assumed that a direct link
between the source and the destination is not available. The half
duplex constraint is accomplished by time sharing between the source
and the relay; i.e.~each transmission period is divided into two time
slots: the source transmits during the first time slot and the relay
during the second one. The relay remains silent during the source
transmission and vice versa. It is assumed that the source does not
have access to any statistical or deterministic channel state
information (CSI). The signal received at the relay
($\boldsymbol{y}_\text{R}$) due to the source transmission
is given by 
\begin{equation}
\label{eq:S-RTransmission}
\boldsymbol{y}_\text{R}=\boldsymbol{H}_1 \boldsymbol{x}+\boldsymbol{w}_\text{R}
\end{equation}
where the $n_\text{R}\times n_\text{S}$ matrix $\boldsymbol{H}_1$
represents channel between the source and the relay
(below, only the statistics of $\boldsymbol{H}_1$ are
assumed to be known at the relay). With $P_\text{S}$ the power
constraint of the source, the column vector $\boldsymbol{x}$ is the
signal transmitted from the source with
$\boldsymbol{Q}=\mathbb{E}(\boldsymbol{x}\boldsymbol{x}^H)=\frac{P_\text{S}}{n_\text{S}}\boldsymbol{I}_{n_\text{S}}$
and the column vector $\boldsymbol{w}_\text{R}$ represents the
receiver noise in the relay with elements independently drawn from a
complex Gaussian random variable with variance $N_0$. The relay
multiplies $\boldsymbol{y}_\text{R}$ with gain matrix $\boldsymbol{F}$
and forwards it to the destination. Then, the received signal at the
destination is 
\begin{eqnarray}
\label{eq:R-DTransmission}
\boldsymbol{y}_\text{D}&=&\boldsymbol{h}_2 \boldsymbol{F}\boldsymbol{y}_\text{R}+\boldsymbol{w}_\text{D}
\\
&=&\boldsymbol{h}_2\boldsymbol{F}\boldsymbol{H}_1 \boldsymbol{x}+\boldsymbol{h}_2\boldsymbol{F}\boldsymbol{w}_\text{R}+\boldsymbol{w}_\text{D}\nonumber
\end{eqnarray}
where the row vector $\boldsymbol{h}_2$ indicates the channel between
the relay and the destination; $\boldsymbol{w}_\text{D}$ represents
the receiver noise at the destination. For simplicity, we assume that
$\boldsymbol{w}_\text{D}$ is statistically equivalent to
$\boldsymbol{w}_\text{R}$ and that both noise processes have
unit-variance, i.e., $N_0=1$; the latter choice is no extra
restriction, as the ratios of transmit powers and noise powers
determine performance, and we are still free to choose $P_\text{S}$
and $P_\text{R}$ arbitrarily.

We assume spatial correlation only at the relay, which can be due to
unobstructed relay node or space limits at the relay
which force antennas to be closely located. Justifications to assume
transceivers with spatial correlation can be found
in \cite{ShFoGaKa:2000, Sh:1999}. The correlation matrix in the relay
is represented by $\boldsymbol{\Sigma}$ with spectral decomposition
$\boldsymbol{\Sigma}
= \boldsymbol{U}_\Sigma\boldsymbol{\Lambda}_\Sigma\boldsymbol{U}^H_\Sigma$,
where $\boldsymbol{U}_\Sigma$ is a unitary matrix with its columns
the eigenvectors corresponding to $\boldsymbol{\Sigma}$, and
$\boldsymbol{\Lambda}_\Sigma$ is diagonal matrix with the eigenvalues of
$\boldsymbol{\Sigma}$ in decreasing order.  Note that we assume
i.i.d. channels for $\boldsymbol{H}_1$ and
$\boldsymbol{h}_2$. Therefore, the channel matrices $\boldsymbol{H}_1$
and $\boldsymbol{h}_2$ can be written using the Kronecker model
as 
\begin{eqnarray}
\label{eq:H--1Cor}
\boldsymbol{H}_1 &=& \boldsymbol{\Sigma}^{\frac{1}{2}}\boldsymbol{H}_{1w}
\\
\label{eq:h--2Cor}
\boldsymbol{h}_2 &=& \boldsymbol{h}_{2w} \boldsymbol{\Sigma}^{\frac{1}{2}}
\end{eqnarray}
where $\boldsymbol{H}_{1w}$ and $\boldsymbol{h}_{2w}$ are i.i.d., zero
mean, unit variance complex Gaussian random variables, independent of
each other; $\boldsymbol{\Sigma}^{\frac{1}{2}}$ denotes a matrix
formed by the square roots of the elements in the matrix
$\boldsymbol{\Sigma}$.

\subsection{Problem Statement}
\label{Subsec:Problem Statement}
With (\ref{eq:R-DTransmission}), the ergodic capacity of the
system is defined as 
\begin{eqnarray}
\label{eq:ErgodicCapactityDefinition}
C= \frac{1}{2} \max_{\substack{\boldsymbol{Q}=\frac{P_\text{S}}{n_\text{S}}\boldsymbol{I} \\F: \mathbb{E}\lbrace\| \boldsymbol{F} \boldsymbol{y}_\text{R} \|^2\rbrace\le P_\text{R}}} \mathbb{E} \lbrace C(\boldsymbol{H}_1,\boldsymbol{h}_2,\boldsymbol{F})\rbrace
\end{eqnarray}
where the relay gain matrix $\boldsymbol{F}$ is to be designed to
maximise (\ref{eq:ErgodicCapactityDefinition}), when the expectation
operation is carried out over $\boldsymbol{H}_1$ and
$\boldsymbol{h}_2$. Assuming
$\boldsymbol{Q}=\frac{P_\text{S}}{n_\text{S}}\boldsymbol{I}_{n_\text{S}}$
(this means equal transmit power from each antenna  is chosen in the source,
because no channel knowledge is available there), the channel
capacity $C(\boldsymbol{H}_1,\boldsymbol{h}_2,\boldsymbol{F})$ for
given channel matrices is 
\begin{eqnarray}
\label{eq:ErgodicCapactityLog}
C(\boldsymbol{H}_1,\boldsymbol{h}_2,\boldsymbol{F}) = \log\big(1+\frac{P_\text{S}}{n_\text{S}}\frac{ \boldsymbol{h}_2\boldsymbol{F}\boldsymbol{H}_1\boldsymbol{H}^H_1\boldsymbol{F}^H\boldsymbol{h}^H_2}{N_0(1+\boldsymbol{h}_2\boldsymbol{F}\boldsymbol{F}^H\boldsymbol{h}^H_2)}\big)
\end{eqnarray}
where
$N_0(1+\boldsymbol{h}_2\boldsymbol{F}\boldsymbol{F}^H\boldsymbol{h}^H_2)$
is the total equivalent noise power which remains constant per
coherence time due to a block fading assumption we impose.

Two major problems are addressed below:
\paragraph*{Problem 1} Solving the optimization problem in
 (\ref{eq:ErgodicCapactityDefinition}), in order to find the optimal
 precoder $\boldsymbol{F}^o$ in the relay that maximizes mutual
 information between the transmit signal from the source and the
 received signal at the destination, given that the correlation
 matrix $\boldsymbol{\Sigma}$ as well as the unit variance of elements  of i.i.d $\boldsymbol{H}_{1w}$ and $\boldsymbol{h}_{2w}$ are available at the
 relay.

\paragraph*{Problem 2} Obtaining necessary and sufficient 
conditions based on $\boldsymbol{\Sigma}$, so that maxium eigenmode
relaying (MER) is optimal.

\section{Optimal Precoding in the Relay}
\label{Sec:Optimal Precoding in the Relay} 
In this section, \textit{Problem 1} from Section \ref{Subsec:Problem
Statement} is addressed. However, before we continue to define
$\boldsymbol{F}^o$, a closer look at the power constraint of the relay
in (\ref{eq:ErgodicCapactityDefinition}) is provided. The power
constraint of the relay is given by 
\begin{eqnarray}
\label{eq:PowConstraintRelay}
 \mathbb{E}\lbrace \| \boldsymbol{F} \boldsymbol{y}_\text{R} \|^2\rbrace= \mathbb{E}\lbrace\mathrm{Tr}(\boldsymbol{F} \boldsymbol{y}_\text{R}\boldsymbol{y}^H_\text{R}\boldsymbol{F}^H)\rbrace
 \le P_\text{R}.
\end{eqnarray} 
One can write
$\mathrm{Tr}(\boldsymbol{F} \boldsymbol{y}_\text{R}\boldsymbol{y}^H_\text{R}\boldsymbol{F}^H)
=\mathrm{Tr}(\boldsymbol{F}^H\boldsymbol{F} \boldsymbol{y}_\text{R}\boldsymbol{y}^H_\text{R})$. Note
that $\boldsymbol{F}$ is a positive gain matrix; therefore, $\boldsymbol{G}
= \boldsymbol{F}^H\boldsymbol{F}$ is a positive symmetric matrix. One
can also choose the gain matrix $\boldsymbol{F}$ to be symmetric (i.e. $\boldsymbol{F} = \boldsymbol{G}^\frac{1}{2}$), which follows from substituting (\ref{eq:H--1Cor}) and (\ref{eq:h--2Cor}) in (\ref{eq:R-DTransmission}) where $\boldsymbol{F}$ is left and right multiplied by $\boldsymbol{\Sigma}^{\frac{1}{2}}$; therefore, by choosing $\boldsymbol{F} = \boldsymbol{G}^\frac{1}{2}$, the problem
of finding $\boldsymbol{F}^o$ will be replaced with finding
$\boldsymbol{G}^o$. By substituting (\ref{eq:S-RTransmission}) and
(\ref{eq:H--1Cor}) in (\ref{eq:PowConstraintRelay}) and applying
spectral decomposition of $\boldsymbol{\Sigma}$,
(\ref{eq:PowConstraintRelay}) will be further simplified
to 
\begin{eqnarray}
\label{eq:RelayPowConstraintG}
 \mathbb{E}\lbrace\mathrm{Tr}(\boldsymbol{G} \boldsymbol{y}_\text{R}\boldsymbol{y}^H_\text{R})\rbrace
=P_\text{S}\mathrm{Tr}(\boldsymbol{G}\boldsymbol{U}_\Sigma\boldsymbol{\Lambda}_\Sigma\boldsymbol{U}^H_\Sigma)
 +N_0 \mathrm{Tr}(\boldsymbol{G})\le P_\text{R} \: .
\end{eqnarray}
 By combining (\ref{eq:RelayPowConstraintG}) and (\ref{eq:ErgodicCapactityDefinition}), we have 
\begin{eqnarray}
\label{eq:ErgodicCapactityDefinition2}
C= \frac{1}{2} \hspace{-4ex} \max_{\substack{\boldsymbol{Q}=\frac{P_\text{S}}{n_\text{S}}\boldsymbol{I} \\\boldsymbol{G}: P_\text{S}\mathrm{Tr}(\boldsymbol{G}\boldsymbol{U}_\Sigma\boldsymbol{\Lambda}_\Sigma\boldsymbol{U}^H_\Sigma)
 +N_0 \mathrm{Tr}(\boldsymbol{G})\le P_\text{R}}}  \hspace{-4ex} \mathbb{E} \lbrace C(\boldsymbol{H}_1,\boldsymbol{h}_2,\boldsymbol{G})\rbrace.
\end{eqnarray}
Assuming (\ref{eq:H--1Cor}) and (\ref{eq:h--2Cor}), applying spectral
decomposition according to $\Sigma
= \boldsymbol{U}_\Sigma\boldsymbol{\Lambda}_\Sigma\boldsymbol{U}^H_\Sigma$
and considering that the statistics of $\boldsymbol{H}_1$ and
$\boldsymbol{h}_2$ do not change by a multiplication with a unitary
matrix, the ``instantaneous'' capacity (\ref{eq:ErgodicCapactityLog})
for given channels $\boldsymbol{H}_1,\boldsymbol{h}_2$ is obtained
as 
\begin{eqnarray}
\label{eq:ErgodicCapactityLogG--hat}
C(\boldsymbol{H}_1,\boldsymbol{h}_2,\hat{\boldsymbol{G}})
= \log\big(1+\frac{\gamma \boldsymbol{h}_{2w}\hat{\boldsymbol{G}}^\frac{1}{2}\boldsymbol{H}_{1w}\boldsymbol{H}^H_{1w}\hat{\boldsymbol{G}}^\frac{1}{2}\boldsymbol{h}^H_{2w}}{n_\text{S}(1+\boldsymbol{h}_{2w}\hat{\boldsymbol{G}}\boldsymbol{h}^H_{2w})}\big)
\end{eqnarray}
where $\gamma = P_\text{S}/N_0$
and 
\begin{equation}
\label{eq:EigenDecompositionG--hat}
\hat{\boldsymbol{G}}^\frac{1}{2} = \boldsymbol{\Lambda}_{\Sigma}^\frac{1}{2} \boldsymbol{U}^H_\Sigma \boldsymbol{G}^\frac{1}{2}\boldsymbol{U}_\Sigma\boldsymbol{\Lambda}_{\Sigma}^\frac{1}{2} \: .
\end{equation}
 Then, the relay power constraint in (\ref{eq:RelayPowConstraintG}) can be written as 
\begin{eqnarray}
\label{eq:RelayPowConstraintG--hat}
P_\text{S}\mathrm{Tr}(\boldsymbol{\Lambda}^{-1}_\Sigma\hat{\boldsymbol{G}})
 +N_0 \mathrm{Tr}(\boldsymbol{\Lambda}^{-2}_\Sigma\hat{\boldsymbol{G}})\le P_\text{R}.
\end{eqnarray}

On the other hand, applying the spectral decomposition
$\hat{\boldsymbol{G}} = \boldsymbol{U}_{\hat{G}} \boldsymbol{\Lambda}_{\hat{G}} \boldsymbol{U}^H_{\hat{G}}$ in (\ref{eq:ErgodicCapactityLogG--hat}) and considering that  the statistics of a random
matrix do not change by multiplying with a unitary matrix, so we
obtain 
\begin{eqnarray}
\label{eq:ErgodicCapactityLogLambdaG--hat}
C(\boldsymbol{H}_1,\boldsymbol{h}_2,\boldsymbol{\Lambda}_{\hat{G}}) = \log\big(1+\frac{\gamma \boldsymbol{h}_{2w}\boldsymbol{\Lambda}_{\hat{G}}^\frac{1}{2}\boldsymbol{H}_{1w}\boldsymbol{H}^H_{1w}\boldsymbol{\Lambda}_{\hat{G}}^\frac{1}{2}\boldsymbol{h}^H_{2w}}{n_\text{S}(1+\boldsymbol{h}_{2w}\boldsymbol{\Lambda}_{\hat{G}}\boldsymbol{h}^H_{2w})}\big) \; .
\end{eqnarray}
By careful inspecting (\ref{eq:ErgodicCapactityLogG--hat}) and (\ref{eq:ErgodicCapactityLogLambdaG--hat}) , the 
conclusion is that 
\begin{equation}
\label{eq:CapacityInequality}
\mathbb{E}(C(\boldsymbol{H}_1,\boldsymbol{h}_2,\hat{\boldsymbol{G}})) = \mathbb{E}( C(\boldsymbol{H}_1,\boldsymbol{h}_2,\boldsymbol{\Lambda}_{\hat{G}})) \: .
\end{equation}
One interpretation of (\ref{eq:CapacityInequality}) is that
$C(\boldsymbol{H}_1,\boldsymbol{h}_2,\hat{\boldsymbol{G}})$ in
(\ref{eq:ErgodicCapactityLogG--hat}) will be maximized by choosing
$\hat{\boldsymbol{G}}$ to be a diagonal matrix,
i.e. $\hat{\boldsymbol{G}} = \boldsymbol{\Lambda}_{\hat{G}}$; that is
equivalent to choosing
$\boldsymbol{U}_{\hat{G}}= \boldsymbol{I}$. Considering that the spectral decomposition of $\boldsymbol{G}$ and  $\hat{\boldsymbol{G}}$ in (\ref{eq:EigenDecompositionG--hat}) are $\boldsymbol{G} =\boldsymbol{U}_G\boldsymbol{\Lambda}_G\boldsymbol{U}_G $
and $\hat{\boldsymbol{G}} =\boldsymbol{U}_{\hat{G}}\boldsymbol{\Lambda}_{\hat{G}}\boldsymbol{U}_{\hat{G}} $, respectively, then choosing $\boldsymbol{U}_{\hat{G}}= \boldsymbol{I}$  in
(\ref{eq:EigenDecompositionG--hat}) dictates that $\boldsymbol{U}_{G}
=\boldsymbol{U}_{\Sigma}$ must hold.

So far, it is proved that
 $\boldsymbol{\Lambda}_{\hat{G}}$ can achieve the same capacity as
 $\hat{\boldsymbol{G}}$. However, for a complete proof, it must be
 made sure that by choosing $\hat{\boldsymbol{G}}
 =\boldsymbol{\Lambda}_{\hat{G}}$ the power constraint of the relay
 defined in (\ref{eq:RelayPowConstraintG--hat}) is
\emph{not} violated. It is clear from (\ref{eq:RelayPowConstraintG--hat}) 
that the power constraint of the relay depends on
 $\hat{\boldsymbol{G}}$ only via two terms:
 $\mathrm{Tr}(\boldsymbol{\Lambda}^{-1}_\Sigma\hat{\boldsymbol{G}})$
 and
 $\mathrm{Tr}(\boldsymbol{\Lambda}^{-2}_\Sigma\hat{\boldsymbol{G}})$. In
 what follows, we will prove that by choosing $\hat{\boldsymbol{G}}
 =\boldsymbol{\Lambda}_{\hat{G}}$, both terms in the relay power
 constraint will be minimized, hence, fulfilling the constraint.

\emph{\textbf{Lemma}: }
Let $\hat{\boldsymbol{G}}$ be an arbitrary positive symmetric matrix
with spectral decomposition $\hat{\boldsymbol{G}}
= \boldsymbol{U}_{\hat{G}} \boldsymbol{\Lambda}_{\hat{G}} \boldsymbol{U}^H_{\hat{G}}
$. Let $\boldsymbol{\Lambda}_{\Sigma}$ be a diagonal matrix with
the elements on the diagonal in descending order and $k>0$.
Then 
\begin{eqnarray}
\label{eq:Lemma1}
\mathrm{Tr}(\boldsymbol{\Lambda}^{-k}_\Sigma\boldsymbol{\Lambda}_{\hat{G}})\le\mathrm{Tr}(\boldsymbol{\Lambda}^{-k}_\Sigma\hat{\boldsymbol{G}}).
\end{eqnarray}
\begin{proof}
See \cite[Theorems 1-3]{JaViGo:2001} for a detailed proof for
$k=1$. The proof can easily be extended to arbitrary $k\geq 0$.
\end{proof}
\noindent
By applying (\ref{eq:Lemma1}) to (\ref{eq:RelayPowConstraintG--hat}), the
power constraint will be fulfilled; that is, by choosing
$\boldsymbol{U}_{G} =\boldsymbol{U}_{\Sigma}$, or equivalently
$\hat{\boldsymbol{G}}=\boldsymbol{\Lambda}_{\hat{G}}= \boldsymbol{\Lambda}_G \boldsymbol{\Lambda}^2_\Sigma$,
it is assured that 
\begin{eqnarray}
\label{eq:RelayPowConstraintLambda}
&&\hspace{-20mm}P_\text{S}\mathrm{Tr}(\boldsymbol{\Lambda}^{-1}_\Sigma\boldsymbol{\Lambda}_{\hat{G}})
 +N_0\mathrm{Tr}(\boldsymbol{\Lambda}^{-2}_\Sigma\boldsymbol{\Lambda}_{\hat{G}})
 \leq \nonumber\\
&&\hspace{5mm}P_\text{S}\mathrm{Tr}(\boldsymbol{\Lambda}^{-1}_\Sigma\hat{\boldsymbol{G}})
 +N_0\mathrm{Tr}(\boldsymbol{\Lambda}^{-2}_\Sigma\hat{\boldsymbol{G}})
 \leq P_\text{R}.
\end{eqnarray}

By completing the proof for the power constraint of the relay, the
answer to \textit{Problem 1} in Section \ref{Sec:System Model and
Problem Statement} is accomplished. Therefore, the transmission from
the relay should be conducted in the direction of the eigenvectors of
the correlation matrix $\boldsymbol{\Sigma}$. However, the power per
eigenvector ($\lambda_i^G$) must be determined using numerical
methods (see e.g.~\cite{JoHo:2003} for a single hop system model).

So far, we assumed equal correlation matrix $\boldsymbol{\Sigma}$ in $\boldsymbol{H}_1$ and $\boldsymbol{h}_2$ which implies that the correlation is a consequence of space limit in the relay node, however, when the correlation occurs due to unobstructed relay node, the correlation matrices corresponding to $\boldsymbol{H}_1$ and $\boldsymbol{h}_2$  can be different. Nevertheless, derivation of $\boldsymbol{F}$ will not be much different. Let assume that $\boldsymbol{H}_1= \boldsymbol{\Sigma}_1^{\frac{1}{2}}\boldsymbol{H}_{1w}$ and $\boldsymbol{h}_2= \boldsymbol{h}_{2w}\boldsymbol{\Sigma}_2^{\frac{1}{2}}$. Then, applying  a similar method as explained in this section, one can prove that the optimal relaying matrix is 
\begin{eqnarray}
\label{eq:unequal}
F=U_{\Sigma_{2}} \Lambda U^H_{\Sigma_{1}}
\end{eqnarray}
where $U_{\Sigma_{i}}$, $i\in \lbrace 1,2 \rbrace$, is a unitary matrix corresponding to the eigenvectors of  $\Sigma_i$ and $\Lambda$ is a diagonal matrix with its elements in decreasing order which is to be calculated numerically.

\section{Optimality of Maximum Eigenmode Relaying}
\label{Sec:Optimality of Maximum Eigenmode Relaying}
In this Section, \textit{Problem 2} from Section \ref{Subsec:Problem Statement} is addressed. The focus of this section is to derive necessary and sufficient
conditions under which MER is the optimal transmission method from the
relay. 

By the results obtained in Section \ref{Sec:Optimal Precoding
in the Relay} and considering that
$\boldsymbol{\Lambda}_{\hat{G}}= \boldsymbol{\Lambda}_G \boldsymbol{\Lambda}^2_\Sigma$,
the ergodic capacity in (\ref{eq:ErgodicCapactityDefinition}) can be
rewritten as 
\begin{eqnarray}
\label{eq:ErgodicCapactityDefinitionLambdaG}
C= \frac{1}{2} \max_{\substack{\boldsymbol{Q}=\frac{P_\text{S}}{n_\text{S}}\boldsymbol{I} \\\boldsymbol{\Lambda}_G:P_\text{S}\mathrm{Tr}(\boldsymbol{\Lambda}_\Sigma\boldsymbol{\Lambda}_{G})
 +N_0\mathrm{Tr}(\boldsymbol{\Lambda}_{G})\le P_\text{R}}} \mathbb{E} \lbrace C(\boldsymbol{H}_{1w},\boldsymbol{h}_{2w},\boldsymbol{\Lambda}_G)\rbrace
\end{eqnarray}
where the expectation-operation is carried out over
$\boldsymbol{H}_{1w}$ and $\boldsymbol{h}_{2w}$
with 
\begin{eqnarray}
\label{eq:ErgodicCapactityLogLambdaG}
&&\hspace*{-20mm}C(\boldsymbol{H}_{1w},\boldsymbol{h}_{2w},\boldsymbol{\Lambda}_G) = \log\big(1+\nonumber\\
&&\frac{\gamma \boldsymbol{h}_{2w}\boldsymbol{\Lambda}_G^\frac{1}{2}\boldsymbol{\Lambda}_\Sigma\boldsymbol{H}_{1w}\boldsymbol{H}^H_{1w}\boldsymbol{\Lambda}_G^\frac{1}{2}\boldsymbol{\Lambda}_\Sigma\boldsymbol{h}^H_{2w}}{n_\text{S}(1+\boldsymbol{h}_{2w}\boldsymbol{\Lambda}_G\boldsymbol{\Lambda}^{2}_\Sigma\boldsymbol{h}^H_{2w})}\big) \; .
\end{eqnarray}
Then $C(\boldsymbol{H}_{1w},\boldsymbol{h}_{2w},\boldsymbol{\Lambda}_G) $ can be simplified according to 
\begin{eqnarray}
\label{eq:ErgodicCapactityLogLambdaG--Sum}
C(\boldsymbol{H}_{1w},\boldsymbol{h}_{2w},\boldsymbol{\Lambda}_G) = \log\big(1+\frac{\gamma\sum\limits_{i=1}^{n_\text{S}} \mid \boldsymbol{h}_{2w}\boldsymbol{\Lambda}_{G}^{\frac{1}{2}}\boldsymbol{\Lambda}_{\Sigma}\boldsymbol{h}_{1w,i}\mid^2}{n_\text{S}(1+ \boldsymbol{h}_{2w}\boldsymbol{\Lambda}_G\boldsymbol{\Lambda}^{2}_\Sigma\boldsymbol{h}^H_{2w})})
\end{eqnarray}
where $\boldsymbol{h}_{1w,i}$ represents the $i$th column of
$\boldsymbol{H}_{1w}$. One can write the
numerator 
\begin{eqnarray}
\label{eq:NumeratorManipulation}
\frac{1}{n_\text{S}}\sum_{i=1}^{n_\text{S}}\mid \boldsymbol{h}_{2w}\boldsymbol{\Lambda}_{G}^{\frac{1}{2}}\boldsymbol{\Lambda}_{\Sigma}\boldsymbol{h}_{1w,i}\mid^2 
=
\sum_{j=1}^{n_\text{R}} {\lambda}_{j}^{G}{\lambda}_{j}^{\Sigma 2}  X_jY
\end{eqnarray}
and the denominator
\begin{eqnarray}
\label{eq:DenomiatorManipulation}
\boldsymbol{h}_{2w}\boldsymbol{\Lambda}_G\boldsymbol{\Lambda}^{2}_\Sigma\boldsymbol{h}^H_{2w} =  \sum_{j=1}^{n_\text{R}}{\lambda}_{j}^{G}{\lambda}_{j}^{\Sigma 2}X_j
\end{eqnarray}
where $X_j= \mid {h}_{2w,j}\mid^2$ is an exponential random variable
with unit mean, i.e. $f_{X_j}(t)= e^{-t}$, and $Y
= \frac{1}{n_\text{S}}\sum_{i=1}^{n_\text{S}} \mid{h}_{1w,i}\mid^2$ is
the sum of $n_\text{S}$ i.i.d exponential random variables with
parameter $n_\text{S}$. Indeed, $Y$ has an Erlang-distribution with
rate and shape equal to $n_\text{S}$, i.e. $f_{Y}(t)=
(t^{n_\text{S}}-1)e^{-t}/({n_\text{S}}-1)!$.  The proof for the
identity (\ref{eq:NumeratorManipulation}) is omitted due to lack of space
but a similar proof can be found
in \cite[Eq. (71)-(72)]{DhMcMaBe:2011}.
Plugging in (\ref{eq:NumeratorManipulation}) and
(\ref{eq:DenomiatorManipulation}) into
(\ref{eq:ErgodicCapactityLogLambdaG--Sum}), it will simplify
to 
\begin{eqnarray}
\label{eq:ErgodicCapactityLogLambdaG--Simple}
&&\hspace{-5mm}C(\boldsymbol{H}_{1w},\boldsymbol{h}_{2w},\boldsymbol{\Lambda}_G) =\log\big(1+\frac{\gamma Y\sum_{j=1}^{n_\text{R}} {\lambda}_{j}^{G}{\lambda}_{j}^{\Sigma 2}  X_j }{1+ \sum_{j=1}^{n_\text{R}} {\lambda}_{j}^{G}{\lambda}_{j}^{\Sigma 2}  X_j }\big)\\
&& \hspace{-5mm}= \log\big(1+\sum_{j=1}^{n_\text{R}} {\lambda}_{j}^{G}{\lambda}_{j}^{\Sigma 2} ( 1 + \gamma Y  )X_j\big)
-\log\big(1+ \sum_{j=1}^{n_\text{R}} {\lambda}_{j}^{G}{\lambda}_{j}^{\Sigma 2}  X_j \big). \nonumber
\end{eqnarray}

\subsection{Necessary Condition}
\label{sub:Necessary Condition}
Now we turn the attention to the question under which condition 
 MER is optimal; i.e. the condition under which $ \lambda_{j}^{G}=
0$ for $j\geq 2$ is the capacity achieving transmission method
with 
\begin{equation}
\label{eq:Lambda1_value}
\lambda_{1}^{G}=  \frac{P_\text{R}}{(1+\lambda^{\Sigma}_1P_\text{S})}
\end{equation}
 which is calculated by substituting $ \lambda_{j}^{G}= 0$ for $j\geq 2$ in (\ref{eq:RelayPowConstraintLambda}).
 
 In order to evaluate the necessary and sufficient condition on
 the optimality of MER,  we assume
 $P=\sum_{j=1}^{n_\text{R}}\lambda^{G}_j$. Assume that the power $P-p$
 is allocated to the dominant eigenvector of $\boldsymbol{G}$,
 i.e. $\lambda^{G}_1=P-p$, and the power $p$ is allocated to the
 remaining eigenvectors of $\boldsymbol{G}$,
 i.e. $p=\sum_{j=2}^{n_\text{R}}\lambda^{G}_j$. It is clear that if
 MER is optimal, then $\partial C(p)/\partial p\mid_{p=0}\leq 0$ which
 provides the necessary condition for the optimality of MER and
 $\partial^2 C(p)/\partial p^2 \mid_{p=0}\leq 0$ which confirms the
 sufficiency. It can be proved that $\partial C(p)/\partial
 p\mid_{p=0}$ will be maximized if $\lambda^{G}_2 = p$ and
 $\lambda^{G}_j = 0$ for $j >2$; similar discussions are provided
 in \cite{DhMcMaBe:2011, JaViGo:2001,ViMa:2001} and, hence, we do not
 repeat it here. Therefore, given the assumption $\lambda^{G}_1 = P-p$
 and $\lambda^{G}_2 = p$,
 (\ref{eq:ErgodicCapactityLogLambdaG--Simple}) will be further
 simplified to 
\begin{eqnarray}
\label{eq:ErgodicCapactityLogLambdaG1and2}
&&\hspace*{-5mm}C(\boldsymbol{H}_{1w},\boldsymbol{h}_{2w},\boldsymbol{\Lambda}_G) =\\
 &&\log\bigg(1+(P-p){\lambda}_{1}^{\Sigma 2}( 1 + \gamma Y) X_1+ p{\lambda}_{2}^{\Sigma 2} ( 1 + \gamma Y)X_2\bigg)\nonumber\\
&&\hspace*{22mm}-\log\bigg(1+ (P-p){\lambda}_{1}^{\Sigma 2}  X_1+p{\lambda}_{2}^{\Sigma 2}  X_2 \bigg)\nonumber
\end{eqnarray}
and so, $\partial C(p)/\partial p\mid_{p=0}$ from (\ref{eq:ErgodicCapactityDefinitionLambdaG}) and  
(\ref{eq:ErgodicCapactityLogLambdaG1and2}) equals 
\begin{eqnarray}
\label{eq:FirstDerivitive}
\frac{\partial}{\partial p} \mathbb{E}\lbrace C(\boldsymbol{H}_{1w},\boldsymbol{h}_{2w},\boldsymbol{\Lambda}_G) \rbrace\mid_{p=0}
&=&\mathbb{E}\lbrace\frac{{\lambda}_{2}^{\Sigma 2} ( 1 + \gamma Y)X_2}{1+P{\lambda}_{1}^{\Sigma 2} ( 1 + \gamma Y)X_1}\rbrace\nonumber\\
&-&\mathbb{E}\lbrace\frac{{\lambda}_{1}^{\Sigma 2}( 1 + \gamma Y)  X_1}{1+P{\lambda}_{1}^{\Sigma 2} (1 + \gamma Y)X_1}\rbrace\nonumber\\
&-&\mathbb{E}\lbrace \frac
{{\lambda}_{2}^{\Sigma 2}  X_2 - {\lambda}_{1}^{\Sigma 2}  X_1}
{1+ P{\lambda}_{1}^{\Sigma 2}  X_1}
\rbrace .
\end{eqnarray}
The first expectation in  (\ref{eq:FirstDerivitive}) can be written as 
\begin{eqnarray}
\label{eq:FirstExpectation}
\mathbb{E}\lbrace\frac{{\lambda}_{2}^{\Sigma 2} ( 1 + \gamma Y)X_2}{1+P{\lambda}_{1}^{\Sigma 2}( 1 + \gamma Y)X_1}\rbrace 
=
{\lambda}_{2}^{\Sigma 2}
\mathbb{E}\lbrace\frac{  ( 1 + \gamma Y)}{1+P{\lambda}_{1}^{\Sigma 2}  ( 1 + \gamma Y)X_1}\rbrace
\end{eqnarray}
which follows from the fact that $X_2$ is independent of $X_1$ and $Y$ and that $\mathbb{E}\lbrace X_2\rbrace = 1$. By simple manipulations, the second  expectation in  (\ref{eq:FirstDerivitive}) equals 
\begin{eqnarray}
\label{eq:SecondExpectation}
\mathbb{E}\lbrace\frac{{\lambda}_{1}^{\Sigma 2} (1  + \gamma Y)X_1}{1+P{\lambda}_{1}^{\Sigma 2} (1  + \gamma Y)X_1}\rbrace 
&=&\\
&&\hspace{-15mm}\frac{1}{P}-\frac{1}{P}
\mathbb{E}\lbrace\frac{1}{1+P{\lambda}_{1}^{\Sigma 2} ( 1 + \gamma Y)X_1}\rbrace\nonumber
\end{eqnarray}
and the last expectation in (\ref{eq:FirstDerivitive}) can be written as 
\begin{eqnarray}
\label{eq:ThirdExpectation}
\mathbb{E}\lbrace \frac
{{\lambda}_{2}^{\Sigma 2}  X_2 - {\lambda}_{1}^{\Sigma 2}  X_1}
{1+ P{\lambda}_{1}^{\Sigma 2}  X_1}
\rbrace
&=&\\ 
&&\hspace{-15mm}-\frac{1}{P}+ 
\frac
{e^{\frac{1}{P{\lambda}_{1}^{\Sigma 2} }}(1+{P{\lambda}_{2}^{\Sigma 2} }) \Gamma(0,\frac{1}{P{\lambda}_{1}^{\Sigma 2} })}
{P^2{\lambda}_{1}^{\Sigma 2}}\nonumber
\end{eqnarray}  
where $\Gamma(0,z)$ is the incomplete Gamma function \cite[Sec. 6.5]{AbSt:2000}. 
By combining (\ref{eq:ErgodicCapactityDefinitionLambdaG}), (\ref{eq:FirstDerivitive}), (\ref{eq:FirstExpectation}),
 (\ref{eq:SecondExpectation}) and (\ref{eq:ThirdExpectation}), in order to compute $\partial C(p)/\partial p \mid_{p=0} \leq 0$, we have
\begin{eqnarray}
\label{eq:FirstDerivativeLeq}
\frac{\partial C(p)}{\partial p} \mid_{p=0} &=& \mathbb{E}\lbrace \frac{{\lambda}_{2}^{\Sigma 2}( 1+\gamma  Y) + 1/P}{1+P{\lambda}_{1}^{\Sigma 2} ( 1+\gamma  Y)X_1} \rbrace \\
&& - \frac
{e^{\frac{1}{P{\lambda}_{1}^{\Sigma 2} }}(1+{P{\lambda}_{2}^{\Sigma 2} }) \Gamma(0,\frac{1}{P{\lambda}_{1}^{\Sigma 2} })}
{(P{\lambda}_{1}^{\Sigma})^2} \leq 0 \nonumber
\end{eqnarray}
from which by elementary operations the following constraint will be
obtained for the MER to be
optimal: 
\begin{eqnarray}
\label{eq:MERConstraintFirst}
{\lambda}_{2}^{\Sigma 2} \leq
\frac{ \mathrm{e}^{\frac{1}{P{\lambda}_{1}^{\Sigma 2} }} \Gamma(0,\frac{1}{P{\lambda}_{1}^{\Sigma 2} }) -P{\lambda}_{1}^{\Sigma 2}\mathbb{E}\lbrace\frac{1}{Z} \rbrace   }
{ P^2{\lambda}_{1}^{\Sigma 2}\mathbb{E}\lbrace \frac{( 1+\gamma  Y)}{Z}\rbrace - P\mathrm{e}^{\frac{1}{P{\lambda}_{1}^{\Sigma 2} }} \Gamma(0,\frac{1}{P{\lambda}_{1}^{\Sigma 2} }) }
\end{eqnarray}
where $Z=1+P{\lambda}_{1}^{\Sigma 2} ( 1+\gamma Y)X_1$.  Note that
$\mathbb{E}\lbrace1/Z \rbrace $ and $\mathbb{E}\lbrace ( 1+\gamma
Y)/Z\rbrace$ in (\ref{eq:MERConstraintFirst}) are non-trivial and do
not seem to have closed-form solutions. However, they can be
calculated either using monte carlo simulations or
by numerical integration of
\begin{eqnarray}
\label{eq:ExpectationNumerical}
\mathbb{E}\lbrace\frac{1}{Z}\rbrace&=&\int_{0}^{\infty} \mathcal{A}(t) \mathrm{e}^{\mathcal{A}(t)} \Gamma(0,\mathcal{A}(t) ) \mathrm{d}t\\
\label{eq:ExpectationNumerical2}
\mathbb{E}\lbrace\frac{1+\gamma  Y}{Z}\rbrace &=&\frac{1}{P{\lambda}_{1}^{\Sigma 2}}\int_{0}^{\infty}  \mathrm{e}^{\mathcal{A}(t)} \Gamma(0,\mathcal{A}(t) ) \mathrm{d}t
\end{eqnarray}
with 
\begin{equation}
\label{eq:A(1)}
\mathcal{A}(t) =  \frac{1 }{P{\lambda}_{1}^{\Sigma 2} ( 1+\gamma  t)} \: .
\end{equation}
The proof of (\ref{eq:ExpectationNumerical}) and
({\ref{eq:ExpectationNumerical2}) follows from the definition of the
expectation operation and is omitted due to lack of space.

\subsection{Sufficient Condition}
\label{sub:Sufficien Condition}
By deriving the necessary condition for MER to be optimal, it can be
proved that the necessary condition is also a sufficient condition for
MER to be optimal. It will be proved by showing that $\partial^2C(p)/\partial p^2 \leq 0$ for arbitrary $p \in[0,P] $. 

For deriving $\partial^2C(p)/\partial p^2$, from (\ref{eq:ErgodicCapactityDefinitionLambdaG}) and  
(\ref{eq:ErgodicCapactityLogLambdaG1and2})
we have 
\begin{eqnarray}
\label{eq:SecondDerivitive}
&&\frac{\partial^2}{\partial p^2} \mathbb{E}\lbrace C(\boldsymbol{H}_{1w},\boldsymbol{h}_{2w},\boldsymbol{\Lambda}_G) \rbrace\mid_{p=0}=\\
&&\hspace*{-7mm}- \frac{P_\text{S}\big((\lambda_1^{\Sigma})^2 X_1  -   (\lambda_2^{\Sigma})^2 X_2\big)^2  \big(  2 +\gamma Y+ 2P(\lambda_1^{\Sigma})^2  X_1(1+\gamma Y) \big )  Y}
{n_\text{S} N_0\big(1+P(\lambda_1^{\Sigma})^2X_1\big)^2   \big(1+P(\lambda_1^{\Sigma})^2X_1(1+\gamma Y)\big)^2}.\nonumber
\end{eqnarray}
It is clear from (\ref{eq:SecondDerivitive}) that $\partial^2C(p)/\partial p^2 \leq 0$ for every $p$, and so, the necessary condition for the optimality of MER derived in (\ref{eq:MERConstraintFirst}) is sufficient as well.

%

\section{Discussion}
\label{Sec:Discussion}
In the previous section, a necessary and sufficient condition were
derived which show the optimality region of MER. As it is clear from
(\ref{eq:FirstDerivativeLeq}) or (\ref{eq:MERConstraintFirst}), there
is need to perform Monte Carlo simulations or numerical integrations
to validate the optimality of MER. However, computationally expensive
Monte Carlo simulations and also numerical integration methods are not
a good choice in practical real-time communication systems; therefore,
in the sequel, we will investigate two simplified approaches which
lead to closed form solutions. 

We first derive a lower bound for the necessary and the sufficient
conditions for the optimality of MER which are a direct result of
Jensen's inequality. Consequently, if the lower bound condition
confirms the optimality of MER, there is no need for the Monte Carlo
simulations and, hence, MER can be used as a capacity-achieving
method.

Then we investigate the effect of the source antenna array on the
optimality of MER. That is done by considering large antenna arrays in
the source using the central limit theorem. Novel, closed form
necessary and sufficient conditions are derived for the optimality of
MER which, according to the simulations, turn out to very tightly
approximate the results for arbitrary numbers of antennas in the
source.

\subsection{A Lower Bound on the Optimality of Maximum Eigenmode Relaying }
\label{Subsec:A Lower bound on the Optimality of Maximum Eigenmode Relaying}
From previous discussions, it is clear that $\partial C(p)/\partial p
<0\mid_{p=0}$ defines the condition under which MER is optimal. We will
exploit it to derive necessary and sufficient conditions which
specify a lower bound on the optimality of MER.  From careful
inspection of (\ref{eq:FirstDerivativeLeq}) it is clear
that 
\begin{eqnarray}
\label{eq:JensenLeq}
\frac{\lambda_2^{\Sigma 2}(1+\gamma Y) + 1/P}{1+P{\lambda}_{1}^{\Sigma 2} ( 1+\gamma  Y)X_1} >0
\end{eqnarray}
which equals the expression inside the expectation operation in
(\ref{eq:FirstDerivativeLeq}). On the other hand, according to
Jensen's inequality  whe have $\mathbb{E}\lbrace
g(X)\rbrace\geq g(\mathbb{E}\lbrace X \rbrace)$ for a convex function
$g(x)$. Let us assume $g(X) = 1/X$, which is a convex function for
$X>0$. Therefore, by applying Jensen's inequality to the expectation
operation in (\ref{eq:FirstDerivativeLeq}), we
have
\begin{eqnarray}
\label{eq:ExpecationLeq}
&&\mathbb{E}\lbrace\frac{\lambda_2^{\Sigma 2}(1+\gamma Y) + 1/P}{1+P{\lambda}_{1}^{\Sigma 2} ( 1+\gamma  Y)X_1} \rbrace> \\
&&\hspace{30mm}1 / \mathbb{E}\lbrace\frac{1+P{\lambda}_{1}^{\Sigma 2} ( 1+\gamma  Y)X_1}{\lambda_2^{\Sigma 2}(1+\gamma Y) + 1/P} \rbrace\nonumber
\end{eqnarray}
with 
\begin{eqnarray}
\label{eq:InvExpectation}
&&\hspace*{-10mm} \mathbb{E}\lbrace\frac{1+P{\lambda}_{1}^{\Sigma 2} ( 1+\gamma  Y)X_1}{\lambda_2^{\Sigma 2}(1+\gamma Y) + 1/P} \rbrace
 =\\
 &&\hspace{6mm}\frac{P\big(\lambda_2^{\Sigma 2}(1+P\lambda_1^{\Sigma 2}) + n_\text{S}(\lambda_1^{\Sigma 2}-\lambda_2^{\Sigma 2})
 \mathrm{e}^{\mathcal{D}}\mathrm{E}_{n_\text{S}+1}(\mathcal{D})\big)}{\lambda_2^{\Sigma 2}(1+P\lambda_2^{\Sigma 2})}\nonumber
\end{eqnarray}
where $\mathcal{D} = n_\text{S}(1+P\lambda_2^{\Sigma 2})/\gamma
P \lambda_2^{\Sigma 2}$ and $\mathrm{E}_m(z)$ the Exponential Integral
function \cite[Sec. 5.1]{AbSt:2000}. Therefore\footnote{The proof of the expectation operation in (\ref{eq:InvExpectation}) is omitted due to space limit, however, it can be validated using symbolic tools, e.g. Mathematica.}, the following
expression is defined as a lower bound which specifies that MER is
optimal: 
\begin{eqnarray}
\label{eq:LowerBound}
&&\hspace{-10mm} \frac{P\lambda_1^{\Sigma 2}\lambda_2^{\Sigma 2}(1+P\lambda_2^{\Sigma 2})}{\lambda_2^{\Sigma 2}(1+P\lambda_1^{\Sigma 2}) + n_\text{S}(\lambda_1^{\Sigma 2}-\lambda_2^{\Sigma 2})
 \mathrm{e}^{\mathcal{D}}\mathrm{E}_{n_\text{S}+1}(\mathcal{D})}
<\\
&&\hspace{30mm}e^{\frac{1}{P{\lambda}_{1}^{\Sigma 2} }}(1+{P{\lambda}_{2}^{\Sigma 2} }) \Gamma(0,\frac{1}{P{\lambda}_{1}^{\Sigma 2} })\nonumber \: .
\end{eqnarray}

\subsection{Large Antenna Array in the Source}
\label{Subsec:Large Antenna Array in the Source}
It is interesting to investigate the effect of the dimension of the
antenna array on the optimality of MER. Note that as we focus on the
optimality of maximum eigenmode transmission from the relay; therefore
we only consider the effect of large antenna arrays on the optimality
of MER.
 \begin{figure}
\begin{center}
        \includegraphics[width=0.465\textwidth]{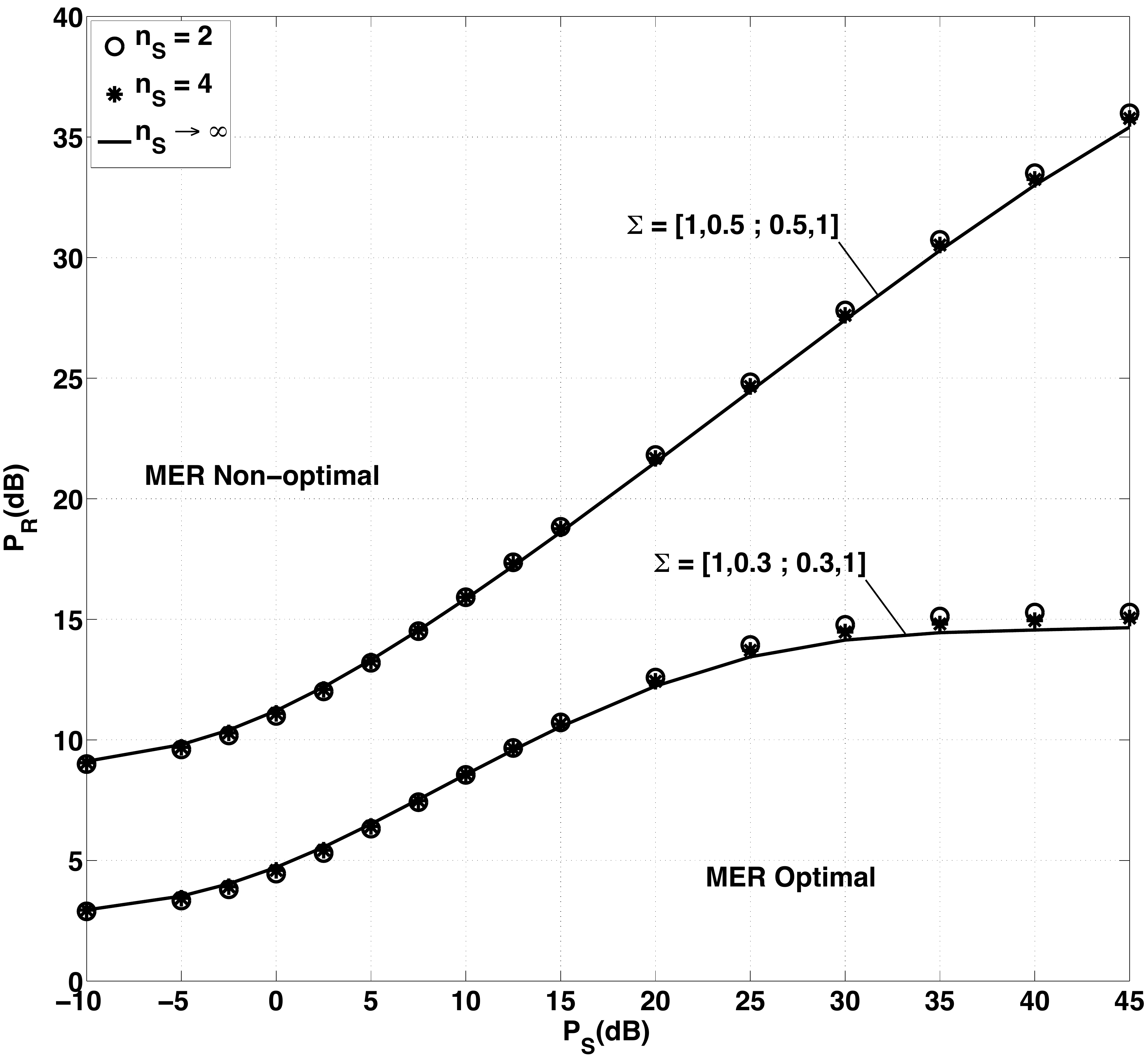}
\end{center} 
\caption{MER optimality region for ($P_\text{S}$,$P_\text{R}$) pairs for 
various correlations $\Sigma$.
 }
\label{fig:Pr-Ps}
\vspace{1.5ex}
\end{figure} 
Assuming $n_\text{S}\gg1$, the random variable $Y \sim
E(n_\text{S},n_\text{S})$ can be approximated by a Gaussian distribution
with mean $1$ and variance $1/n_\text{S}$, i.e., approximately,
$Y \sim \mathcal{N}(1,1/n_\text{S}$). Therefore, as
$n_\text{S} \to \infty$, we obtain $Y\to 1$. By substituting $Y=1$ in
(\ref{eq:MERConstraintFirst}), we have $1+\gamma Y = 1+\gamma
$. Hence, $\mathbb{E}\lbrace1/Z\rbrace$ in
(\ref{eq:MERConstraintFirst}) can be written in closed form
as 
\begin{eqnarray}
\label{eq:FirstExpectation--NsInf}
\mathbb{E}\lbrace\frac{1}{Z}\rbrace 
=
\mathcal{A}_1 \mathrm{e}^{\mathcal{A}_1}\Gamma(0,\mathcal{A}_1)
\end{eqnarray}
where $\mathcal{A}_1 = 1 /P{\lambda}_{1}^{\Sigma 2} (1+\gamma)$ is
obtained by substituting $t=1$ in (\ref{eq:A(1)}). Then, the necessary
and sufficient conditions for the optimality of MER, i.e. $\partial
C(p)/\partial p \mid_{p=0}\leq 0$ in (\ref{eq:FirstDerivativeLeq}) or
(\ref{eq:MERConstraintFirst}) for large $n_{\text{S}}$, after some
manipulation lead to the following closed-form constraint:
\begin{eqnarray}
\label{eq:MERConstraint--LargeNs}
{\lambda}_{2}^{\Sigma 2} \leq
\frac{ \mathrm{e}^{\frac{1}{P{\lambda}_{1}^{\Sigma 2} }} \Gamma(0,\frac{1}{P{\lambda}_{1}^{\Sigma 2} }) -P{\lambda}_{1}^{\Sigma 2}\mathcal{A}_1 \mathrm{e}^{\mathcal{A}_1}\Gamma(0,\mathcal{A}_1) \rbrace }
{ P\mathrm{e}^{\mathcal{A}_1}\Gamma(0,\mathcal{A}_1) - P\mathrm{e}^{\frac{1}{P{\lambda}_{1}^{\Sigma 2} }} \Gamma(0,\frac{1}{P{\lambda}_{1}^{\Sigma 2} }) } \: .
\end{eqnarray}
It is clear from (\ref{eq:MERConstraint--LargeNs}) that with large
antenna array in the source, the optimality of MER depends on the source
transmission only via $P_\text{S}$ and it is independent of
$\boldsymbol{H}_1$. 

 It will be shown, by numerical simulations in the next Section, that the
closed-form constraint in (\ref{eq:MERConstraint--LargeNs})
approximates (\ref{eq:MERConstraintFirst}) with high accuracy.

\begin{figure}
\begin{center}
        \includegraphics[width=0.465\textwidth, height =0.43 \textwidth]{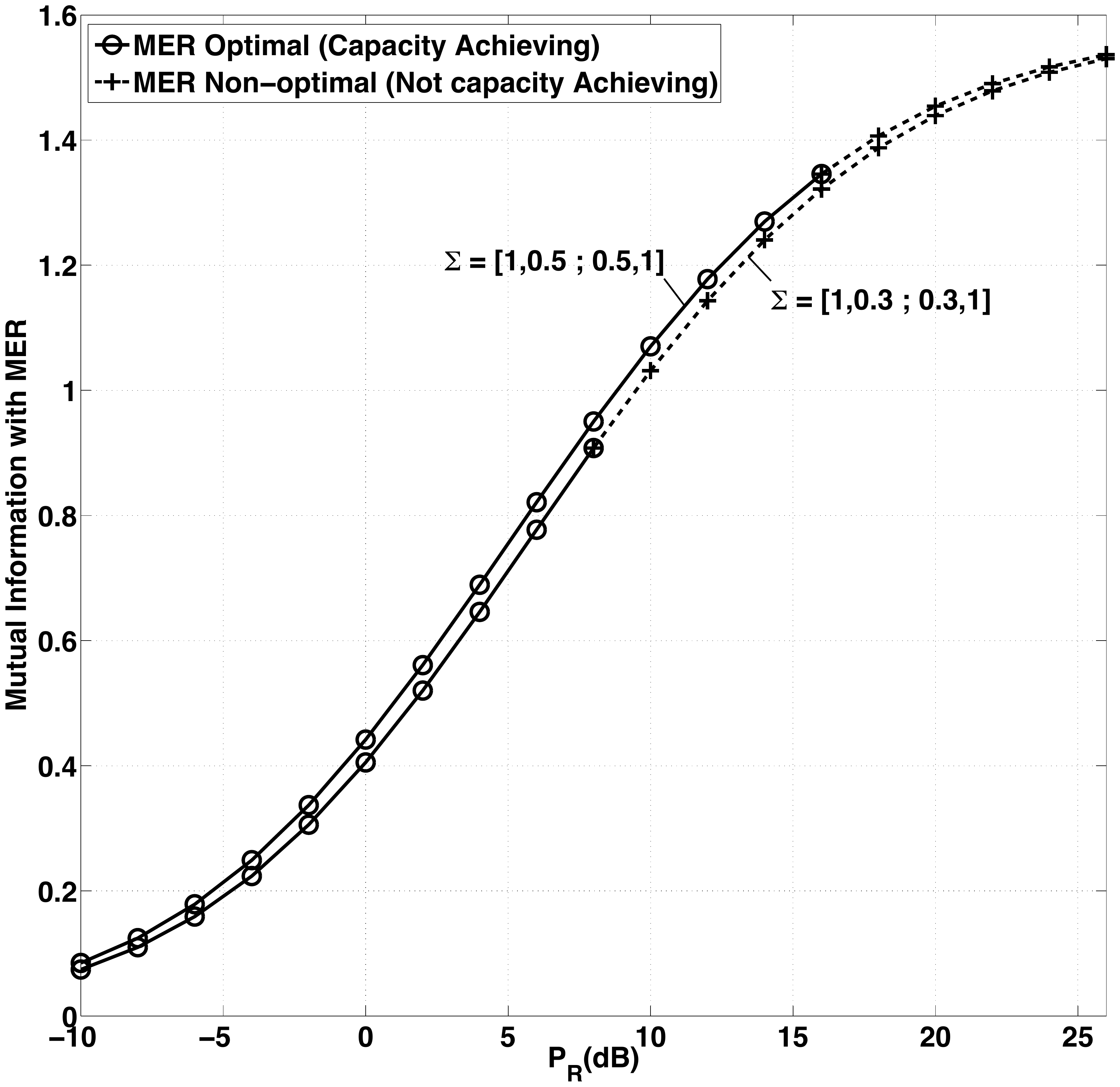}
\end{center} 
\caption{Mutual information between input and output of the relay system 
with $n_\text{S}=2=n_\text{R}$ and MER. Given
 $P_\text{S}=10 \mathrm{dB}$, the $P_\text{R}$ corresponding to solid
 lines illustrates the region where MER achieves capacity, while dashed
 lines show the region where MER can not achieve capacity.  }
\label{fig:C-Pr}
\vspace{1.5ex}
\end{figure} 

 \section{Numerical Results}
\label{Sec:Numerical Results}
Computer simulations are provided to verify the analytical results
derived in the previous sections. We assume that the relay is equipped
with two antennas, but various numbers of the antennas in the source
are evaluated. As explained in Section \ref{Sec:System Model and
Problem Statement}, we assume i.i.d $\boldsymbol{H}_{1,w}$ and
$\boldsymbol{h}_{2,w}$, where the elements are circularly symmetric
complex Gaussian random variables with zero mean and unit variance
(block Rayleigh fading assumption). The noise power in the relay and
the destination is assumed to be $N_0 =1$. Fig. \ref{fig:Pr-Ps}
illustrates the $(P_\text{S},P_\text{R})$-pairs for which MER is
optimal. The optimality region of MER is calculated for $\rho = 0.3$
and $0.5$ where $\rho$ is defined as the inter-antenna correlation or
the off-diagonal elements of the correlation matrix
$\boldsymbol{\Sigma}$. It is clear from Fig.~\ref{fig:Pr-Ps} that by
increasing $\rho$, the optimality region of MER increases,
too. Another interesting observation from Fig.~\ref{fig:Pr-Ps} is that
the optimality of MER is \emph{almost} independent of the number of
antennas in the source node. The figure illustrates the optimality
region using the expression derived in (\ref{eq:MERConstraintFirst})
for $n_\text{S} = 2$ and $4$ which require numerical integrations and
also using the closed form expression derived in
(\ref{eq:MERConstraint--LargeNs}) when $n_\text{S}\to \infty$; it
explicitly shows that regardless of $n_\text{S}$, the optimality
regions of MER coincide with very little
difference. Fig. \ref{fig:C-Pr} shows the mutual information between
the input and output of the specified MIMO relay channel. Note that
the solid lines achieve capacity using MER but the dashed lines do not
achieve capacity and, hence, MER is not optimal in this part of the
curve.

\section{Conclusion}
\label{Sec:Conclusion}
A dual hop cooperative system was investigated in this paper. The source and the relay nodes are equipped with multiple antennas and the destination with single antenna. Optimal precoding matrix in the relay was derived. It was shown that the
optimal transmission from the relay should be conducted in the
direction of the eigenvectors of the transmit-channel covariance
matrix. Then, a necessary and sufficient condition was derived, under
which, the relay transmission only from the strongest eigenvector 
achieves capacity; this method of transmission was called Maximum Eigenmode Relaying.
The exact result contains two integrals which need to be solved
numerically. Moreover, a closed form lower bound was derived for the
optimality region of MER. We further investigate to evaluate the
effect of the source antenna array on the optimality of MER and
derived closed form expression when the source is equipped with
infinite antennas. The simulation results show that MER optimal
region with infinite antennas in the source coincides with MER
optimal region when the source is equipped with much lower number of
antennas with inappreciable difference.

\balance
\bibliography{RefMER.bib}

\end{document}